\documentclass[aip,graphicx]{revtex4-1}

\usepackage{geometry}
\usepackage{setspace}
\usepackage{fullpage}
\usepackage{color}
\usepackage{hyperref}
\usepackage{multirow}
\usepackage[pdftex]{graphicx}
\draft 

\begin{document}

\title{Status of Iron Based Superconductors: characteristics and relevant properties for applications} 

\author{Kazumasa\,Iida}\email{iida.kazumasa@nihon-u.ac.jp}
\affiliation{Department of Electric and Electronic Engineering, Nihon University, Narashino 275-8575, Japan and  JST CREST, Kawaguchi 332-0012, Japan}

\date{\today}

\begin{abstract}
Since the discovery of iron-based superconductors (IBSs) on LaFePO in 2006, many types of IBSs have been fabricated. IBSs have usually been compared to cuprates and MgB$_2$, and the methodology of research developed by them have been implemented to IBSs. As a result, many similarities between IBSs and cuprates have been revealed, e.g., the parent compounds being antiferromagnets and grain boundaries being weak-links to some extent. On the other hands, the distinct features of IBSs are highlighted as multiband superconductors (i.e., the 5 bands of Fe 3$d$ orbital crossing Fermi level) and extended $s$--wave symmetry. Additionally, some of the IBSs are topological superconductors that can be possible platforms for quantum computing. In this paper, an overview of IBS research and development in the last 18 years will be reported, involving characteristics of IBSs as well as strategies of increasing the superconducting transition temperature and critical current density. 
\end{abstract}

\maketitle

\section{Introduction}
In 2006, a superconductor with a transition temperature ($T_\mathrm{c}$) of approximately 6\,K was reported in LaOFeP, marking the first iron-based superconductor IBS \cite{ref1}. Initially, it did not attract much attention due to its low $T_\mathrm{c}$. The discovery was surprising, as the material contains iron--an element typically associated with magnetism rather than superconductivity. However, it was not entirely unexpected, as pure iron loses its magnetism when subjected to pressures exceeding 10\,GPa \cite{ref2}. Further increasing the pressure beyond 15\,GPa induces superconductivity with a maximum $T_\mathrm{c}$ of 2\,K observed around 20\,GPa. Interestingly, a superconducting dome forms as a function of pressure.

In 2008, a superconductor was discovered in La(O,F)FeAs \cite{ref3} with a $T_\mathrm{c}$ of 26\,K, marking a significant breakthrough in the scientific community. This material exhibits a superconducting dome that depends on the fluorine content, similar to what is observed in cuprates upon oxygen doping.

Since the discovery of LaFeAs(O,F), numerous IBSs have been reported. These IBSs are commonly referred to using numerical labels corresponding to their chemical composition, such as La-1111 and Ba-122 (BaFe$_2$As$_2$). For all IBSs, the FeAs and Fe$Ch$ (where $Ch$ represents chalcogen elements) tetrahedrons are common structural features that play a crucial role in superconductivity. The $Ln$-1111 family (where $Ln$ represents lanthanoid elements) exhibits a maximum $T_\mathrm{c}$ of 55\,K under ambient pressure\cite{ref4}, which is the highest $T_\mathrm{c}$ among bulk IBSs. In monolayer form, FeSe shows a $T_\mathrm{c}$ of 65\,K \cite{ref5}, which will be discussed later.

In this article, we review IBSs and explore methods for tuning their superconducting properties. Numerous comprehensive review articles on this subject have already been published\cite{ref6, ref7, ref8, ref9}, to which interested readers are encouraged to refer. The structure of this article is as follows: first, we summarize the fundamental properties of IBSs and provide a comparison with cuprates. We then discuss the tuning of $T_\mathrm{c}$ and the critical current density ($J_\mathrm{c}$) through various approaches. Finally, we briefly review recent progress toward practical applications.

\section{Fundamental Properties of IBSs and Their Comparison with Cuprates}
The calculated density of states for parent compound LaFeAsO indicates that the Fe 3$d$-orbitals dominate the total density of states near the Fermi energy\cite{ref10, ref11}. The Fermi surface consists of an electron-like pocket near the M point and a hole-like pocket near the $\Gamma$ point in momentum space, a feature that is consistent across most IBSs.

The magnetic phase diagrams of typical IBSs show that their parent compounds, such as BaFe$_2$As$_2$\cite{ref12}, LaFeAsO \cite{ref13}, and FeTe (11)\cite{ref14}, exhibit antiferromagnetic ordering, similar to cuprates. However, a key distinction lies in their electronic nature: while IBSs are semimetals, cuprates are Mott insulators.

The Ba-122 series includes hole-doped, electron-doped, and isovalent-doped variants. In all cases, the N\'{e}el temperature decreases with increasing doping concentration, and superconductivity emerges beyond a critical doping level. An intriguing aspect of these compounds is the coexistence of superconductivity and antiferromagnetism, which stands in contrast to the behavior observed in La-1111. With increasing doping concentration, the antiferromagnetic phase eventually vanishes, and $T_\mathrm{c}$ reaches a maximum. For P and Co doping, superconductivity is also suppressed at higher doping levels. Conversely, in the case of K doping, the fully substituted KFe$_2$As$_2$ exhibits superconductivity with a $T_\mathrm{c}$ of about 3 K. The $Ln$-1111 and 11 systems also exhibit a superconducting dome, but the $Ln$-1111 system can only be doped up to approximately 20\% when fluorine is the dopant. A major distinction from cuprates is that direct doping of the layer responsible for electrical conduction (FeAs in these compounds, compared to CuO$_2$ in cuprates) does not negatively impact superconductivity.

The temperature dependence of the upper critical field ($H_\mathrm{c2}$) in representative IBS thin films (Fe(Se,Te)\cite{ref15}, Co-doped Ba-122\cite{ref16}, and SmFeAs(O,H)\cite{ref17}) demonstrates that the $H_\mathrm{c2}$ exceeds 50\,T at low temperatures for both major crystallographic orientations. Notably, the $H_\mathrm{c2}$ of SmFeAs(O,H) for $H$$\parallel$$ab$ has been experimentally determined to reach approximately 120\,T at 2.2\,K. The $H_\mathrm{c2}$ anisotropy ($\gamma_\mathrm{Hc2}$) at $T_\mathrm{c}$ is approximately 1 for Fe(Se,Te) and Co-doped Ba-122, and around 1.5 for SmFeAs(O,H). These values exhibit temperature-dependent variations due to the multi-band nature of these compounds.

The comparison between cuprates and IBSs is summarized in Table\,\ref{tab:table1}. Both systems exhibit a high degree of flexibility in material design, as evidenced by the synthesis of numerous compounds. In both systems, the parent compounds are antiferromagnetic; however, while cuprates are insulators, IBSs are classified as semimetals. The superconducting gap symmetry is $d$--wave in cuprates, whereas it is extended $s$--wave in most IBSs, where five bands cross the Fermi surface. This multi-band characteristic is why IBSs are often referred to as multi-band superconductors. It is also noteworthy that the pairing symmetry of K-doped Ba-122 changes from the nodeless $s\pm$--wave to nodal $s$--wave via $s+is$ with increasing K content\cite{ref18}. The pairing mechanism in cuprates and IBSs may involve spin fluctuations. However, for IBSs, orbital fluctuations may also contribute to the pairing mechanism.

The highest $T_\mathrm{c}$ for hole-doped cuprates is 164\,K (zero resistance at 154\,K \cite{ref19} under 15\,GPa pressure), observed in HgBa$_2$Ca$_2$Cu$_3$O$_{8+\delta}$ (Hg-1223) under 20--30\,GPa of pressure \cite{refChu}, whereas in electron-doped systems, $T_\mathrm{c}$ reaches only 30\,K in (La$_{1-x}$Ce$_x$)$_2$CuO$_4$\cite{ref20}. In contrast, the highest $T_\mathrm{c}$ for IBSs is 55\,K in bulk SmFeAs(O,F) compounds\cite{ref4} and 65\,K in monolayers of FeSe \cite{ref5}, both of which are electron-doped. For hole-doped IBS systems, $T_\mathrm{c}$ reaches 38\,K \cite{ref21}, which is significantly lower compared to hole-doped cuprates.

For YBa$_2$Cu$_3$O$_{7-\delta}$ (Y-123), the experimentally evaluated $H_\mathrm{c2}$ for $H$$\parallel$$ab$ was around 240\,T at 5\,K \cite{refHc2-1}, whereas for $H$$\parallel$$c$, it was approximately 120\,T at 4.2\,K \cite{refHc2-2}. Consequently, $\gamma_\mathrm{Hc2}$$\sim$2 for Y-123 at low temperatures. Due to the Pauli limiting effect, this value is slightly lower than the value extracted from the effective mass anisotropy ($\gamma_\mathrm{m}=m_c/m_{ab}$), where ${m_c}$ and $m_{ab}$ are the effective masses along the crystallographic $c$-axis and ${ab}$--plane, respectively, by applying Blatter scaling to the angular dependence of $J_\mathrm{c}$ data \cite{ref22}. This scaling has also been applied to Fe(Se,Te)\cite{scaling-1}, Co-doped Ba122 \cite{scaling-2}, and $Ln$-1111 ($Ln$=La and Nd) \cite{scaling-3, scaling-4}.

\begin{table}
\caption{Comparison between IBSs and cuprates.}
\begin{tabular}{|l|l|l|}
\hline
 & Cuprates & IBSs\\
 \hline
Design flexibility & High & High\\
\hline
Parent compound & AFM, Mott insulator & AFM, semimetal\\
\hline
Gap symmetry & $d$-wave & Extended $s$-wave\\
\hline
\multirow{2}{*}{Maximum $T_\mathrm{c}$} & 164\,K (hole) & 38\,K (hole)\\ 
                                                                    & 30\,K (electron) & 55\,K (electron)\\ \hline
\multirow{3}{*}{$\gamma_\mathrm{Hc2}$ at $T_\mathrm{c}$} &  & 1: Fe(Se,Te) \\ 
                                                                        & $\sim$2: Y-123 & 1.2: Co-doped Ba-122 \\
                                                                        &  & 1.5: SmFeAs(O,H)\\ \hline
\multirow{4}{*}{Mass anisotropy} &  & 2$\sim$3.5: Fe(Se,Te)\\
						  &  $\sim$5--7: Y-123 & 1.5$\sim$2: Co-doped Ba-122\\
						  &  & 1.4$\sim$2.2: NdFeAs(O,F)\\
                                                                    &  & 3$\sim$4.5: LaFeAs(O,F)\\ \hline                                                                    
\end{tabular}
\label{tab:table 1}
\end{table}

\section{Tuning $T_\mathrm{c}$}
As previously stated, Hg-1223 exhibited the highest $T_\mathrm{c}$ of 164\,K when subjected to a pressure of 20--30\,GPa. Similarly, isotropic pressure enhances the $T_\mathrm{c}$ of superconducting FeSe and La-1111. For FeSe, isostatic pressure increases $T_\mathrm{c}$ from 9\,K to 37\,K at a pressure of 7\,GPa \cite{ref24}. However, further increasing the pressure at 13.9\,GPa reduces $T_\mathrm{c}$ below its original value.
For the LaO$_{1-x}$F$_x$FeAs with $x$=0.11, $T_\mathrm{c}$ exhibits a dome-like dependence on pressure, reaching a maximum $T_\mathrm{c}$ of 43\,K at 4\,GPa \cite{ref25}. In the La-1111 system, chemical pressure induced by substituting La with other lanthanoids also increases $T_\mathrm{c}$, achieving a maximum $T_\mathrm{c}$ of 55\,K, as mentioned earlier \cite{ref4, ref26}.

Isostatic and chemical pressure also induce superconductivity in the parent Ba-122 (or Sr-122), with a maximum $T_\mathrm{c}$ of around 30\,K achieved in both cases \cite{ref27, ref28}. Based on these finding, we attempted to induce superconductivity in the parent Ba-122 by applying epitaxial strain. To transfer the strain, an Fe buffer layer was deposited on MgAl$_2$O$_4$(001) substrate, followed by Ba-122 with varying thicknesses\cite{ref29}. The sample with a thickness of less than 30\,nm exhibits an onset $T_\mathrm{c}$ of around 30\,K. Increasing thickness led to strain relaxation, resulting in the absence of superconductivity.

Recently, a more sophisticated approach, known as the superlattice method, has been reported to induce superconductivity. This method used a superlattice of Ba-122 and SrTiO$_3$, controlling the strain by varying the thickness of Ba-122 \cite{ref30}. By adjusting the Ba-122 thickness to around 7\,nm, an Fe--As--Fe bond angle of approximately 110$^\circ$ was obtained, which is close to the ideal tetrahedral configuration. At this bond angle, the highest onset $T_\mathrm{c}$ of 30\,K was achieved.

In 2012, another significant development occurred in our community. FeSe, which has a bulk $T_\mathrm{c}$ of only 9\,K, exhibited a $T_\mathrm{c}$ of 65\,K  in monolayer form, as confirmed by the opening of a superconducting gap \cite{ref5, ref31, ref32}. An angle-resolved photoemission spectroscopy (ARPES) study revealed that in the monolayer, only electron-like pockets were present at the M point. Interestingly, in two-monolayer samples, a hole-like pocket appeared at the $\Gamma$ point, accompanied by a reduction in $T_\mathrm{c}$. Therefore, the electronic structure of the monolayer, specifically the presence of only electron-like pockets, is crucial for achieving high $T_\mathrm{c}$. However, the monolayer is highly sensitive to exposure to air.

To replicate the electronic structure of the FeSe monolayer in a material that remains stable under atmospheric conditions, single crystals of (Li,Fe)OHFeSe, known as the "five-one" compound, have been fabricated by using the ion exchange method \cite{ref33}. ARPES studies have shown that the electronic structure of these crystals is quite similar to that of the FeSe monolayer. However, it is important to note that the $T_\mathrm{c}$ is 42\,K, which is lower than that of the monolayer. Epitaxial thin films of this compound were also fabricated using matrix-assisted hydrothermal epitaxial growth \cite{ref34}, with $T_\mathrm{c}$ remaining the same as the bulk value.

A similar electronic structure to that of the FeSe monolayer was recently achieved in FeSe bulk single crystals through protonation \cite{ref35}. This method involves the electrolysis of water contained in an ionic liquid. When a bias voltage of 3\,V was applied, the H$_2$O is decomposed into protons and O$^{2-}$, allowing the protons to be intercalated into FeSe. The Hall effect measurements confirmed that dominant charge carriers are electrons. The $T_\mathrm{c}$ increased stepwise, reaching up to 45\,K, and notably, this process was completely reversible. However, $T_\mathrm{c}$ is still lower than that of the FeSe monolayer.

To tune the superconducting properties, an electronic double-layer transistor (EDLT) has been employed \cite{ref36}. Unlike traditional field-effect transistors (FETs), which use oxide thin films like SiO$_2$ as gate insulating layers, EDLTs utilize ionic liquids. When a gate voltage is applied, a single molecular layer less than 1\,nm thick acts as the insulating film, allowing nearly 100 times more carriers to be doped into the material compared to a conventional FET. A positive bias voltage results in electron doping, while a negative bias voltage leads to hole doping. The $T_\mathrm{c}$ of thin FeSe ($\sim$0.6\,nm) and Fe(Se,Te) (less than 10\,nm) can be increased using an EDLT, with a positive bias voltage indicating electron doping. For FeSe, $T_\mathrm{c}$ was raised to 35--40\,K \cite{ref37, ref38}, and for Fe(Se,Te), it increased to about 38\,K \cite{ref39}. Notably, this process is also completely reversible.

Unfortunately, the enhancement of $T_\mathrm{c}$ using EDLT is effective only for FeSe and Fe(Se,Te). Despite several attempts on TlFe$_{1.6}$Se$_2$ \cite{ref40}, NdFeAsO, and BaFe$_2$(As,P)$_2$ \cite{ref41}, no other materials have shown a significant increase in $T_\mathrm{c}$. Only changes in resistivity or slight shifts in $T_\mathrm{c}$ were observed. Nevertheless, interesting phenomena were observed in BaFe$_2$(As,P)$_2$, such as the ambipolar suppression of $T_\mathrm{c}$ \cite{ref41}.

All attempts to enhance the $T_\mathrm{c}$ of FeSe or Fe(Se,Te) have been limited to 45\,K, as confirmed by transport and magnetic measurements, which is significantly lower than the 65\,K reported by ARPES. This discrepancy may be attributed to superconducting fluctuations. Further investigation using the Nernst effect to probe these fluctuations is necessary.

\section{Grain boundaries}
\begin{figure}[htbp]
\centering
\includegraphics[width=2.8in]{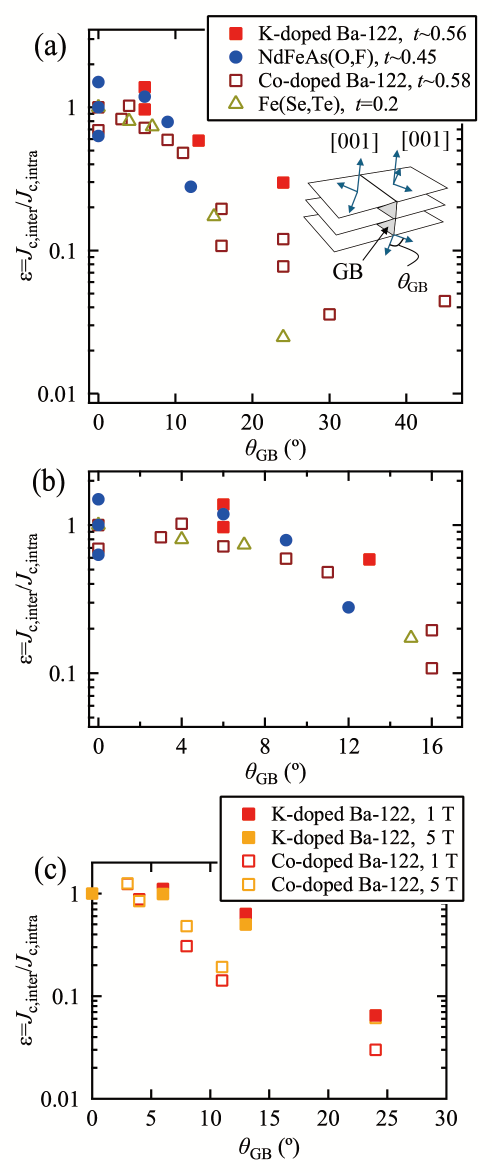}
\caption{(a) The GB transparency $\epsilon$ of K-doped Ba-122 \cite{ref42}, NdFeAs(O,F) \cite{ref43}, Co-doped Ba-122 \cite{ref44}, and Fe(Se,Te) \cite{ref45, ref46}  as a function of misorientation angle ($\theta_\mathrm{GB}$) for the [001]-tilt grain boundary. The measurement temperatures are given in reduced temperatures. (b) A magnified view of (a) for 0$^\circ$$\leq$$\theta_\mathrm{GB}$$\leq$16$^\circ$. (c) A comparison of GB transparency between K-doped Ba-122 (12\,K) and Co-doped Ba-122 (4\,K) in magnetic fields.} \label{fig:Figure1}
\end{figure}

The ratio of inter- to intra-grain $J_\mathrm{c}$, known as grain boundary (GB) transparency ($\epsilon=J_\mathrm{c}^\mathrm{inter}/J_\mathrm{c}^\mathrm{intra}$), for the [001]-tilt GB as a function of misorientation angle for various IBSs is summarized in fig.\,\ref{fig:Figure1}. 
For all IBSs, the critical angle ($\theta_\mathrm{c}$) for which the inter-grain $J_\mathrm{c}$ starts to decrease exponentially,  is around 9$^\circ$ [fig.\,\ref{fig:Figure1}(b)]. This angle is larger than that of cuprates. Another distinct feature is that the inter-grain $J_\mathrm{c}$ remains constant at GB angles larger than 15$^\circ$. Recently, we highlighted the excellent GB properties of K-doped Ba-122 \cite{ref42}. Unlike Co-doped Ba-122, the $\theta_\mathrm{c}$ for K-doped Ba-122 remains unchanged even under an applied magnetic field [fig.\,\ref{fig:Figure1}(c)]. Scanning precision electron diffraction (SPED) revealed that the crystal orientation of MgO bi-crystal substrate changes abruptly at the GB, whereas in K-doped Ba-122, the crystal orientation changes gradually, with an average interval angle of 2$^\circ$. This implies that the GB angle of K-doped Ba-122 is different from that of the substrate, which account for the excellent GB properties observed in K-doped Ba-122.

\begin{figure}[htbp]
\centering
\includegraphics[width=3in]{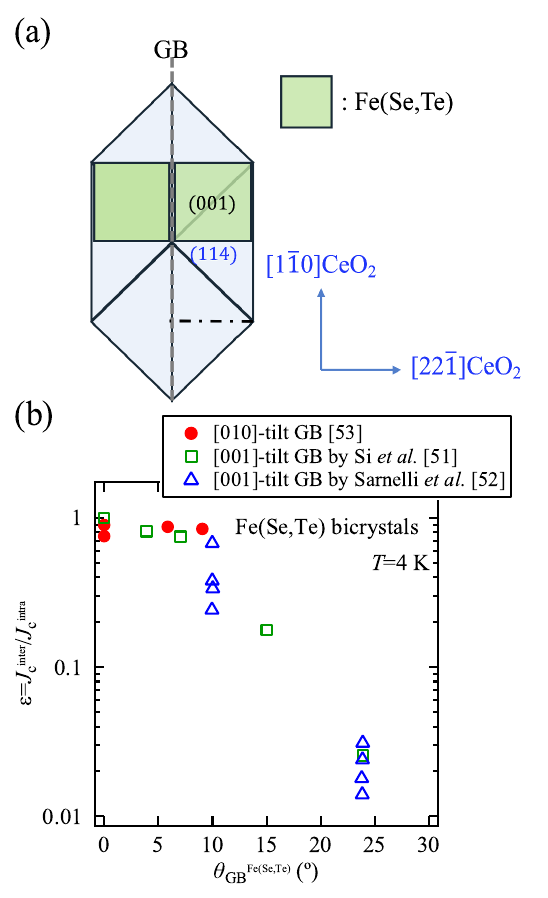}
\caption{(a) Schematic illustration of the out-of-plane for Fe(Se,Te) and CeO$_2$. (b) The ratio of inter- to intra-grain $J_\mathrm{c}$ of Fe(Se,Te) [010]-tilt GB as a function of misorientation angles of Fe(Se,Te) ($\theta_\mathrm{GB}^\mathrm{Fe(Se,Te)}$) \cite{ref47}. For comparison, the data for the [001]-tilt GB are shown \cite{ref45, ref46}.} \label{fig:Figure2}
\end{figure}

As previously mentioned, the [001]-tilt GBs have been investigated for the 11, 122, and 1111 systems. However, other types of GBs in IBSs have yet to be explored. To address this gap, [010]-tilt, roof-type GBs in Fe(Se,Te) were fabricated using PLD \cite{ref47}. To minimize the lattice mismatch between Fe(Se,Te) and the SrTiO$_3$ (STO) bicrystal substrate, a CeO$_2$ buffer layer was deposited. Structural characterization conducted via transmission electron microscope (TEM) and X-ray diffraction (XRD) revealed that the $\theta_\mathrm{GB}$ in Fe(Se,Te) was smaller, whereas the $\theta_\mathrm{GB}$ in CeO$_2$ was larger than that of the substrate. For $\theta_\mathrm{GB}$ of STO exceeding 24$^\circ$, no GB was formed in Fe(Se,Te), whereas the $\theta_\mathrm{GB}$ in CeO$_2$ continued to increase. The inclined growth can be explained by the geometrical coherence model. For GB angles larger than 24$^\circ$, a $\Sigma9[110]/\lbrace221\rbrace$ GB was formed in CeO$_2$. In this case, the epitaxial relationship between Fe(Se,Te) and CeO$_2$ was (001)[100]Fe(Se,Te)$\parallel$(114)[22$\bar{1}$]CeO$_2$. The out-of-plane configuration of Fe(Se,Te) is illustrated in fig.\,\ref{fig:Figure2}(a). Although no data are available beyond a GB angle of 10$^\circ$, the ratio of inter- to intra-grain $J_\mathrm{c}$ for Fe(Se,Te) [001]-tilt GBs is summarized in fig.\,\ref{fig:Figure2}(b). As illustrated, no weak-link behavior was observed.

To increase the inter-grain $J_\mathrm{c}$ for K-doped Ba-122 tapes, the K content was increased to the over-doped regime \cite{ref48}. The intra-grain $J_\mathrm{c}$ peaks at around $x$=0.3 in Ba$_{1-x}$K$_x$Fe$_2$As$_2$, while the maximum inter-grain $J_\mathrm{c}$ was obtained at around $x$=0.48. This is because the over-doped grains enhance the proximity effect.

\section{Tuning $J_\mathrm{c}$}
IBS thin films, especially the 122 systems contain numerous natural defects, such as BaFeO nanopillars \cite{ref49}, stacking faults \cite{ref50}, and dislocations \cite{ref51}, depending on the processing conditions. Notably, these defects are anisotropic, aligning specifically along the $c$-axis, except for stacking fault. This alignment is the reason $J_\mathrm{c}$ for $H$$\parallel$$c$ is higher than that for $H$$\parallel$$ab$.

\begin{figure*}[t]
\centering
\includegraphics[width=6in]{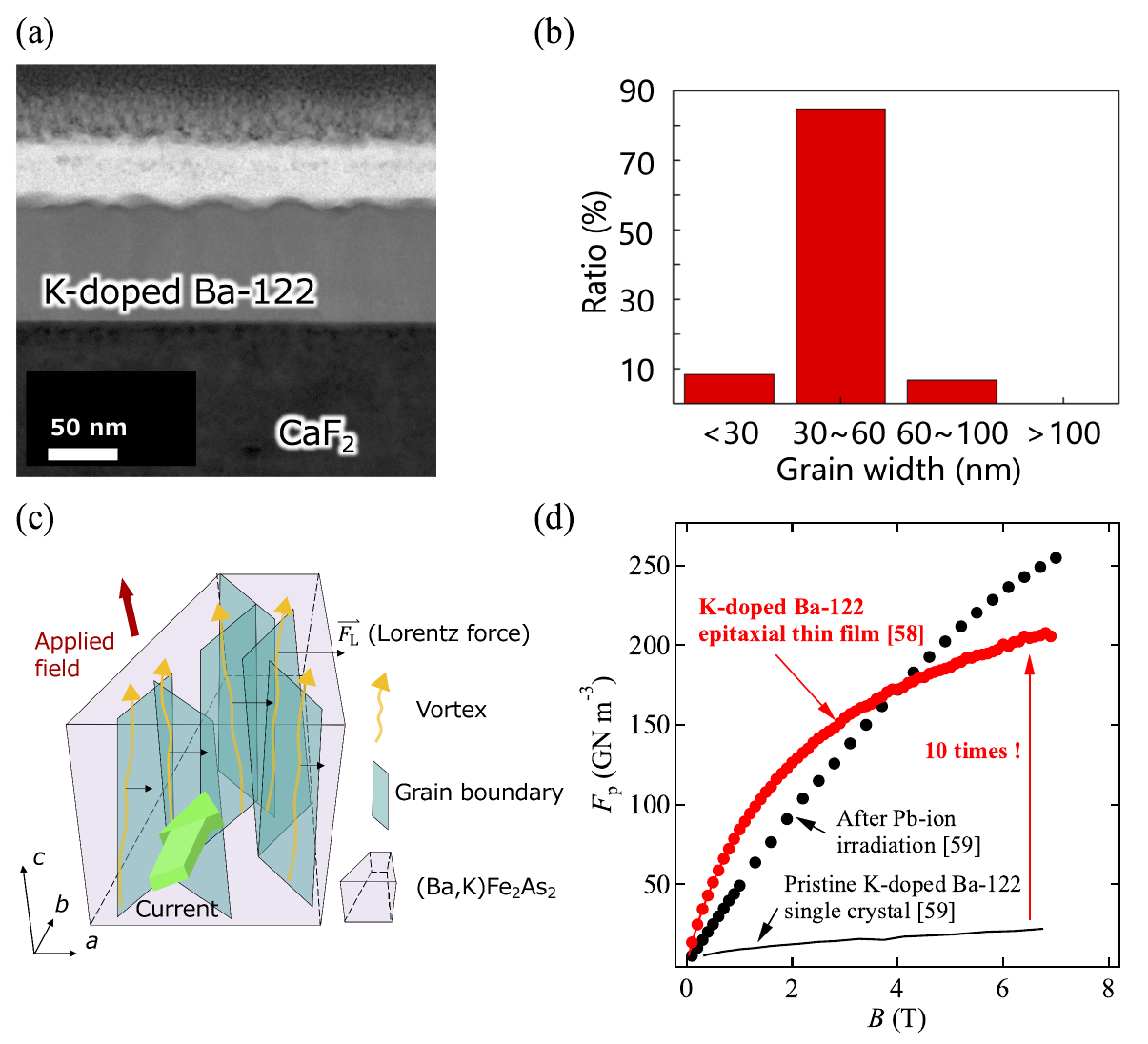}
\caption{(a) Cross-sectional view of the K-doped Ba-122 grown on CaF$_2$ thin film obtained by annular dark field (ADF) scanning transmission electron microscopy (STEM). (b) The distribution of grain width of K-doped Ba-122. (c) A schematic illustration showing how vortices (yellow arrows) are pinned by low-angle grain boundaries in a K-doped Ba-122 superconductor. Black arrows indicate the Lorentz force acting on the vortices. (d) The pinning force density $F_\mathrm{p}$ of K-doped Ba-122 thin film fabricated in this study  is almost 10 times as high as a pristine single crystal \cite{ref53}.} \label{fig:Figure3}
\end{figure*}

Recently, we successfully fabricated epitaxial K-doped Ba-122 thin films using molecular beam epitaxy \cite{ref52}. The key to this success was using fluoride substrates and maintaining a low growth temperature. Under these conditions, columnar grains having a grain width 30$\sim$60 \,nm [fig.\,\ref{fig:Figure3}(b)] grew almost perpendicular to the substrate, with each grain rotating a few degrees [fig.\,\ref{fig:Figure3}(a)]. This created a network of low-angle grain boundaries that act as strong pinning centers, as schematically illustrated in fig.\,\ref{fig:Figure3}(c). As a result, the pinning force density ($F_\mathrm{p}$) of our film is nearly ten times higher than that of single crystals and almost comparable to single crystals irradiated with heavy ions [fig.\,\ref{fig:Figure3}(d)].

To improve the in-field $J_\mathrm{c}$ properties of various IBSs, artificial pinning centers have been introduced by using techniques like the multilayer approach, pulsed laser deposition (PLD) target modification, irradiation, and transition metal doping, as shown in Table\,\ref{tab:table2}. Some of the methods presented in this table are effective in enhancing in-field $J_\mathrm{c}$, which will be summarized later. Irradiation experiments indicate that IBSs are quite robust against disorder. However, in Ba-122, $T_\mathrm{c}$ decreases at a rate of 1\,K per mol\% of BaZrO$_3$ or BaHfO$_3$, which is relatively significant compared to RE-123.

\begin{table}[b]
\caption{Artificial pinning centers in various IBSs. \label{tab:table2}}
\centering
\begin{tabular}{|l|l|l|}
\hline
Materials& Methods & Refs\\ \hline
Co-doped Ba-122 & Multilayer, SrTiO$_3$ insertion layer & \cite{ref54}\\ \hline
Co-doped Ba-122 & Multilayer, Ba-122 insertion layer & \cite{ref55}\\ \hline
Fe(Se,Te) & Multilayer, CeO$_2$ insertion layer & \cite{ref56}\\ \hline
P-doped Ba-122 & BaZrO$_3$ added PLD target & \cite{ref57}\\ \hline
Co-doped Ba-122 & BaZrO$_3$ added PLD target & \cite{ref58}\\ \hline
Co-doped Ba-122 & BaHfO$_3$ added PLD target & \cite{ref59}\\ \hline
FeSe & SrTiO$_3$ added PLD target & \cite{ref60}\\ \hline
Fe(Se,Te) & Proton irradiation & \cite{ref61}\\ \hline
NdFeAs(O,F) & $\alpha$-particle irradiation & \cite{ref62}\\ \hline
(Li,Fe)OHFeSe & Mn doping & \cite{ref63}\\ \hline
\end{tabular}
\end{table}

(Li,Fe)OHFeSe exhibits significant broadening of resistivity curves with increasing magnetic field \cite{refDong-Li}. This behavior is attributed to the large distance of 
9\,\AA\ between FeAs conduction layers. Interestingly, its magnetic phase diagram closely resembles that of Bi-2223. After doping a small amount of Mn into the compound, a significant reduction in resistance broadening was achieved \cite{ref63}. More importantly, the $F_\mathrm{p}$ showed a substantial improvement due to the enhanced in-field $J_\mathrm{c}$, despite $T_\mathrm{c}$ being lowered by 5\,K. Additionally, the residual resistivity ratio decreased from 42 to 13, indicating that considerable amount of disorder was introduced into the material, likely responsible for the improved pinning.

\begin{figure}[t]
\centering
\includegraphics[width=88mm]{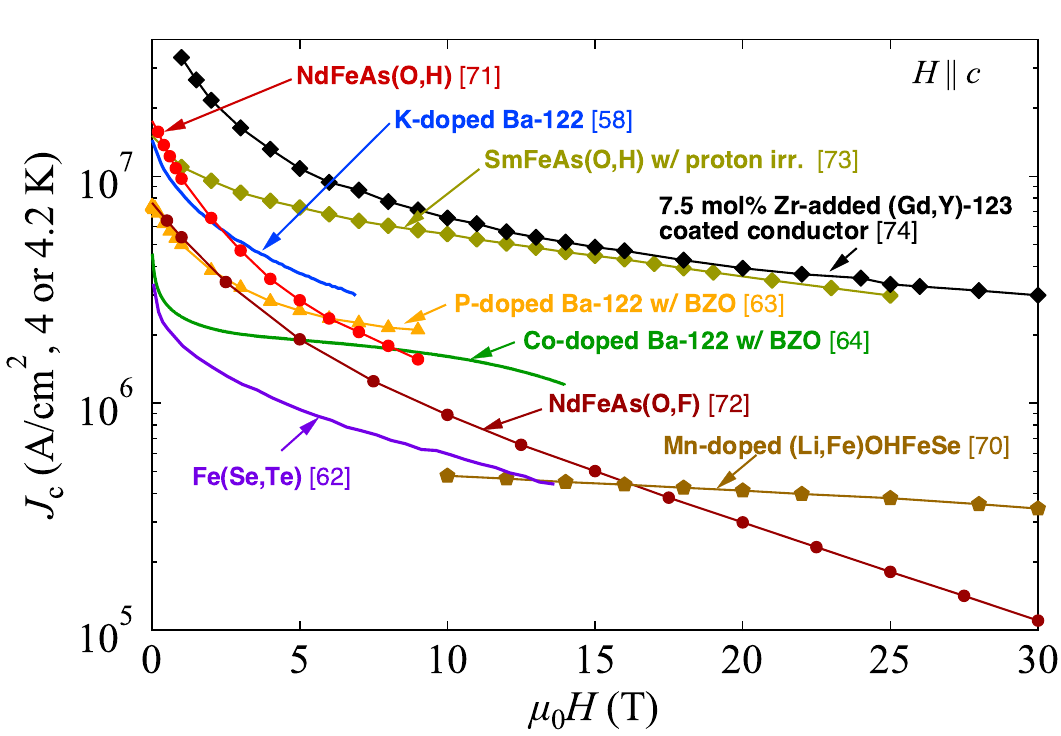}
\caption{The best performance of the $J_\mathrm{c}$--$H$ properties for various IBS thin films measured at 4 or 4.2\,K. Applied magnetic field is parallel to the crystallographic $c$-axis. Solid lines and solid lines with data point represent the $J_\mathrm{c}$ evaluated from the magnetization and transport measurements, respectively. For comparison, 7.5\,mol.\,\% Zr-doped (Gd,Y)-123 coated conductor also superimposed \cite{ref67}.} \label{fig:fig4}
\end{figure}

Finally, the current status of the best-performing $J_\mathrm{c}$--$H$ properties in various IBS thin films grown on single crystals is summarized in fig.\,\ref{fig:fig4}. It is important to note that the self-field $J_\mathrm{c}$ for all IBSs exceed 3\,MA/cm$^2$. This includes Fe(Se,Te)/CeO$_2$ quasi-multilayer, which has a $T_\mathrm{c}$ of 20\,K \cite{ref56}. During the preparation, CeO$_2$ target was ablated with only two pulses. Therefore, neither the CeO$_2$ layer nor particles were observed in the Fe(Se,Te) layer by TEM. However, strained layers were present as confirmed by geometric phase analysis (GPA). These strained layers worked as pinning centers.

Notably, the highest self-field $J_\mathrm{c}$ exceeded 17\,MA/cm$^2$, achieved with NdFeAs(O,H) \cite{ref64}. Although $T_\mathrm{c}$ of both H-doped and F-doped NdFeAsO is around 45\,K, the $J_\mathrm{c}$ of NdFeAs(O,H) was nearly an order of magnitude higher than that of NdFeAs(O,F)\cite{ref65}. This enhancement is attributed to the improvement in the depairing current density ($J_\mathrm{d}$), achieved through heavy electron doping, as the self-field $J_\mathrm{c}$ is almost proportional to $J_\mathrm{d}$, which in turn is proportional to the condensation energy. This thermodynamic approach, combined with the introduction of artificial pinning centers via low-energy proton irradiation, significantly improves $J_\mathrm{c}$--$H$, as evidenced in SmFeAs(O,H) \cite{ref66}. Remarkably, this improvement brings the $J_\mathrm{c}$ values close to that of 7.5\,mol.\,\% Zr-doped (Gd,Y)-123 coated conductor \cite{ref67}, although this data may not be the most recent.

The target modification (i.e., adding BaZrO$_3$ to the PLD target) also improves in-field performance for both P- and Co-doped Ba-122 thin films. For Co-doped Ba-122, BaZrO$_3$ cylinders with a diameter of 4\,nm were embedded in the matrix \cite{ref58}, whereas nano-sized BaZrO$_3$ particles were found in P-doped Ba-122 \cite{ref57}. The reason for the different morphologies of BaZrO$_3$ remains unclear; however, the differences may be influenced by the specific systems and laser sources used.

One of the most remarkable findings was the Mn-doped five-one compound, which maintained an almost constant $J_\mathrm{c}$ of around 0.4\,MA/cm$^2$, even in magnetic fields up to 30\,T \cite{ref63}.

\section{Progress towards applications}
As stated before, IBSs have an anisotropic $s$-wave pairing symmetry, and it was anticipated that the suppression of $J_\mathrm{c}$ at GBs, commonly referred to as the weak-link problem, would be less severe compared to cuprates. Indeed, investigations into the GB properties of various IBSs have revealed that the $\theta_\mathrm{c}$ is approximately 9$^\circ$, which is more than twice as large as that of cuprates. Furthermore, for [001]-tilt grain boundaries, it has been confirmed that the inter-grain $J_\mathrm{c}$ becomes constant when the misorientation angle exceeds 15$^\circ$. Additionally, GBs exhibit metallic characteristics, and the product of resistance in the normal state ($R_\mathrm{n}$) and the GB cross-sectional area ($A$), $R_\mathrm{n}A$, shows values an order of magnitude lower than those of Y-123 \cite{ref675}. These findings indicate that the GB properties of IBSs are superior to those of cuprates.

From an application perspective, the 11, 122, and 1111 families, which are particularly significant, exhibit $H_\mathrm{c2}$ exceeding 50 T at low temperatures -- even in the 11 family, which has the lowest $T_\mathrm{c}$. Additionally, as shown in Table\,\ref{tab:table1}, the anisotropy of $H_\mathrm{c2}$ is relatively low. Moreover, in thin films, it is mentioned in Section V that small-angle GBs can act as flux pinning centers.

Leveraging these characteristics, efforts are underway to develop wires and tapes using the powder-in-tube (PIT) method, as well as polycrystalline bulk materials. The current status of each of these developments is summarized below.

\subsection{Wires and Tapes}
In 2023, a significant advancement was made by a group at Shanghai Jiatong University, where they produced a 1-meter-long Fe(Se,Te) coated conductor \cite{ref68}. This was accompolished using a reel-to-reel system operating at a frequency of 20 to 60\,Hz, with the film having a thickness of 320\,nm. For this 1-meter-long tape, the end-to-end critical current ($I_\mathrm{c}$) was measured at 108\,A, with a $T_\mathrm{c}$ of 17.5\,K. In short samples, the self-field $I_\mathrm{c}$ reached around 175\,A, corresponding to a $J_\mathrm{c}$ of 2.3\,MA/cm$^2$ at 4.2\,K. Additionally, they fabricated a single pancake coil using the Fe(Se,Te) coated conductor \cite{ref69}.

In 2016 the Institute of Electrical Engineering, Chinese Academy of Sciences (IEECAS) has successfully fabricated a 100-meter-long, 7-core K-doped Sr-122 superconducting tape \cite{ref100mlong}.After development and design, these tapes exhibited a uniform $J_\mathrm{c}$ along its entire length, reaching a value of 65\,kA/cm$^2$ at 4.2\,K under an applied magnetic field of 10\,T \cite{ref70}. This uniformity represents a significant milestone in the fabrication of high-performance superconducting tapes. Building on this success, IEECAS manufactured double pancake coils (DPCs) using these tapes, resulting in a total of nine DPCs, from which seven coils with superior critical current $I_\mathrm{c}$ characteristics were selected. These coils were subsequently assembled into high-field insert coils, with each coil connected using Y-123 coated conductor joined by a low melting alloy. During charge testing, the assembled coils achieved a remarkable magnetic field of 21\,T, in combination with a background field of 20\,T. This accomplishment represents the world's first realization of a Tesla-class coil using IBSs.

\begin{figure}[t]
\centering
\includegraphics[width=88mm]{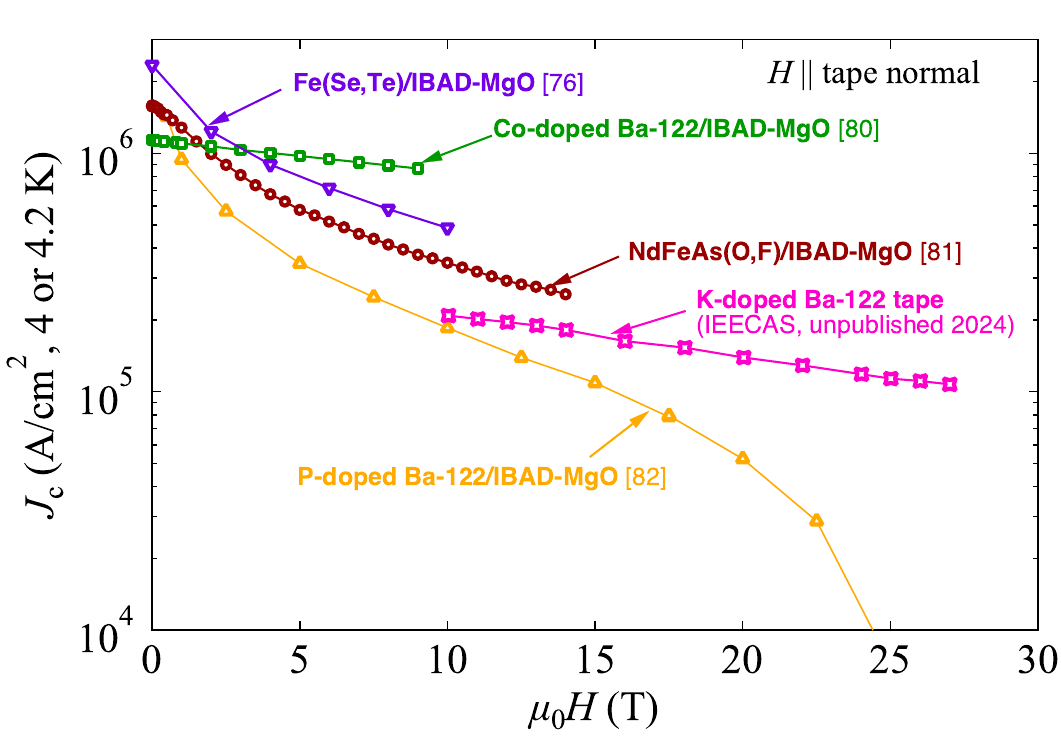}
\caption{The best performance of the $J_\mathrm{c}$--$H$ properties for various IBS tapes measured at 4 or 4.2\,K. Applied magnetic field is parallel to the tape normal. The K-doped Ba-122 tape was fabricated by IEECAS. The data are courtesy by Prof. Yanwei Ma.} \label{fig:fig5}
\end{figure}

Figure\,\ref{fig:fig5} exhibits the current status of $J_\mathrm{c}$--$H$ properties for various IBS. Except for the K-doped Ba-122 tape, all samples were thin films grown on IBAD-MgO, including Fe(Se,Te) \cite{ref68}, Co-doped Ba-122 \cite{ref71}, NdFeAs(O,F) \cite{ref72}, and P-doped Ba-122 \cite{ref73}. Notably, Fe(Se,Te)/IBAD-MgO demonstrated the highest $J_\mathrm{c}$ in the low-field regime despite having the lowest $T_\mathrm{c}$. Additionally, the in-field $J_\mathrm{c}$ of the K-doped Ba-122 tape was superior to that of the biaxially textured P-doped Ba-122 films. This highlights significant advancements in optimizing $J_\mathrm{c}$ for K-doped Ba-122 tapes, which are essential for conductor applications in high-field environments.

\subsection{Bulks}
Yamamoto $et$ $al$. reported on K-doped Ba-122 bulk samples fabricated via both data-driven and researcher-driven process design approaches \cite{ref74}. In the data-driven approach powered by machine learning, a specialized software called BOXVIA \cite{ref75} was employed to optimize the synthesis of K-doped Ba-122 bulk samples. Conversely, the researcher-driven approach utilized human expertise and theoretical models to determine the synthesis parameters. Notably, both approaches contributed to a shared database, facilitating continuous refinement of process parameters. The $J_\mathrm{c}$ as a function of applied magnetic field was compared between these bulk samples. The researcher-driven sample exhibited higher $J_\mathrm{c}$ in the low-field regime, while the data-driven sample demonstrated superior performance under high-field conditions. Micro-structural analysis revealed that the researcher-driven sample contained fine grains of approximately 20-30\,nm, whereas the data-driven sample exhibited a bimodal distribution of grain sizes.

For iron-based superconducting bulk materials, the 1144 compounds are promising candidates for magnet applications. This compound is a hybrid material between two 122 structures and exhibits a $T_\mathrm{c}$ of around 36\,K without any doping \cite{ref76}. The $H_\mathrm{c2}$ exceeds 50\,T, and its anisotropy decreasing from 2.7 to 1.5 with decreasing temperature \cite{Hc2-1144}. Recently, a textured 1144 compound was successfully fabricated using spark plasma sintering (SPS) \cite{ref77}. The $J_\mathrm{c}$--$B$ properties of this textured bulk were found to be superior to those of an untextured reference SPS bulk sample and were comparable to those of K-doped Ba-122.

\subsection{Cost and safety issues}
Regarding costs, it is necessary to consider both material expenses and those associated with safety measures. Kametani, Tarantini, and Hellstrom have recently presented an intriguing report on the production cost of K-doped Ba-122 \cite{ref78}. According to their findings, the raw material cost for K-doped Ba-122 is approximately \$400 per liter. This estimate is based on the small-scale purchase price of raw materials for laboratory use. It is anticipated that further cost reductions could be achieved through large-scale conductor production. Even with a P factor (the ratio of final product cost to raw material cost) of 10, K-doped Ba-122 would cost approximately \$4,000 per liter as a conductor--roughly one-fourth the current cost of Nb$_3$Sn.

There is limited information regarding the adverse health effects on researchers caused by exposure to IBSs. However, GaAs is classified as a Group 1 carcinogen by the International Agency for Research on Cancer (IARC) and is recognized as a hazardous substance \cite{ref79}. Consequently, IBSs may need to be handled in compliance with the safety guidelines established for GaAs. It is also necessary to account for equipment and disposal costs associated with safety measures. Regarding equipment costs, during thin-film fabrication, an arsenic (As) filter must be installed at the exhaust outlet of the vacuum pump to prevent the release of As into the external environment. Additionally, in workplaces where dust is generated (e.g., for PIT wires and bulks), sealed equipment, machinery, or local exhaust ventilation systems must be used at all times. Nevertheless, at this stage, it remains challenging to accurately estimate these costs.

\section{Conclusion and perspective}
A comprehensive review of the current status of iron-based superconductors has been conducted, including an evaluation of various techniques for tuning $T_\mathrm{c}$. It has been observed that high-angle grain boundaries impede supercurrent flow, although not as severely as in cuprates. This characteristic makes IBSs promising candidates for magnet applications involving polycrystalline wires and bulk materials. Furthermore, a strategy to enhance the performance of polycrystalline tapes and bulk materials of K-doped Ba122 has been proposed. The $J_\mathrm{c}$--$B$ performance has been significantly improved through the incorporation of artificial pinning centers (APCs) and a thermodynamic approach combined with APCs, particularly in thin films. Additionally, substantial progress has been made in the development of long-length wires and tapes.

IBSs are capable of operating effectively under magnetic fields below 20\,T at 4.2\,K, or below 10\,T at temperatures ranging from 20 to 30\,K. According to the  perspective by Jaroszynski \cite{ref80}, IBSs have the potential to be competitive with Nb$_3$Sn in terms of both cost and performance. However, numerous challenges remain to be addressed before this potential can be fully realized.

\section*{Acknowledgments}
I would like to express my sincere gratitude to Chiara Tarantini, Jens H\"{a}nisch, Takafumi Hatano, Hiroshi Ikuta, Satoshi Hata, Yusuke Shimada, Akinori Yamanaka, and Akiyasu Yamamoto for their invaluable contributions and collaborative efforts throughout the course of our shared project. I also would like to express my gratitude to the ASC2024, Conference Chair Luisa Chiesa, and Program Co-Chairs Tiina Salmi and Helene Felice for nominating me for the plenary talk. I extend special thanks to the Materials Committee Chair, Ania Kario, for encouraging me to deliver this plenary talk. This work was supported by JST CREST Grant Number JPMJCR18J4 and the JSPS Grant-in-Aid for Scientific Research (B) Grant number 20H02681. This work was also partly supported by the Advanced Characterization Platform of the Nanotechnology Platform Japan sponsored by the Ministry of Education Culture, Sports, Science and Technology (MEXT), Japan.

\end{document}